\documentclass[a4paper,11pt,english]{article}
\usepackage{a4wide}
\usepackage{babel}
\usepackage{amssymb,amsmath}
\usepackage{multicol,color,longtable}
\usepackage[small,bf,hang]{caption}
\usepackage[linktocpage]{hyperref}
\usepackage{array}
\IfFileExists{diagbox.sty}{
	\usepackage{diagbox}
}{
	\newcommand{\diagbox}[2]{##1 \ensuremath{\backslash} ##2}
}

\usepackage[
	backref=true,
	sorting=none,
	maxnames=99,
	firstinits=true,
	doi=false,
	backend=biber,
	citestyle=numeric-comp
	]{biblatex}

\newcolumntype{C}[1]{>{\centering\let\newline\\\arraybackslash\hspace{0pt}}m{#1}}

\newcommand{\be}{\begin{equation}}
\newcommand{\ee}{\end{equation}}

\newcommand{\nn}{\nonumber}
\newcommand{\lp}{\left(}
\newcommand{\rp}{\right)}

\newcommand{\OS}{\mathcal{S}}
\newcommand{\OG}{\mathcal{G}}

\newcommand{\bg}{\bar{g}}

\newcommand{\bcd}{\bar{\nabla}}

\DeclareMathOperator{\sgn}{sgn}

\def\ddal{\mathop{\vrule height 7pt depth0.2pt
\hbox{\vrule height 0.5pt depth0.2pt width 6.2pt}\vrule height 7pt depth0.2pt width0.8pt
\kern-7.4pt\hbox{\vrule height 7pt depth-6.7pt width 7.pt}}}
\def\sdal{\mathop{\kern0.1pt\vrule height 4.9pt depth0.15pt
\hbox{\vrule height 0.3pt depth0.15pt width 4.6pt}\vrule height 4.9pt depth0.15pt width0.7pt
\kern-5.7pt\hbox{\vrule height 4.9pt depth-4.7pt width 5.3pt}}}
\def\ssdal{\mathop{\kern0.1pt\vrule height 3.8pt depth0.1pt width0.2pt
\hbox{\vrule height 0.3pt depth0.1pt width 3.6pt}\vrule height 3.8pt depth0.1pt width0.5pt
\kern-4.4pt\hbox{\vrule height 4pt depth-3.9pt width 4.2pt}}}

\def\dal{\mspace{1.5mu}\mathchoice{\ddal}{\ddal}{\sdal}{\ssdal}\mspace{1.5mu}}

\begin{document}

\pagestyle{empty}
\hfill AEI-2012-060
\vskip 0.1\textheight
\begin{center}
{\Large{\bfseries 
	ON UNITARY SUBSECTORS OF\\[1.3ex]
	 POLYCRITICAL GRAVITIES
}}
\vskip 0.1\textheight

{\bfseries Axel Kleinschmidt}${}^{*\diamond}$, 
{\bfseries Teake Nutma}${}^*$,
{\bfseries Amitabh Virmani}${}^*$ \\%

\vskip 0.5cm
$*$ \\
\emph{\href{http://www.aei.mpg.de/}{Max-Planck-Institut f\"ur
Gravitationsphysik}}\\
\emph{(Albert Einstein Institut)}\\ 
\emph{Am M\"uhlenberg 1, } \\ 
\emph{14476 Golm, Germany}	\\ 
\vskip 0.3cm
$\diamond$ \\
\emph{\href{http://www.solvayinstitutes.be/}{International
Solvay Institutes}}\\
\emph{Campus Plaine C.P. 231,} \\
\emph{Boulevard du Triomphe,} \\
\emph{1050 Bruxelles, Belgium} \\

\vskip 0.5cm
{\tt \{axel.kleinschmidt, teake.nutma,
amitabh.virmani\}@aei.mpg.de}%

\end{center}

\vskip 0.05\textheight

\begin{center} {\bf Abstract } 
\end{center}
\begin{quotation}\noindent
We study higher-derivative gravity theories in arbitrary space-time dimension
$d$ with a cosmological constant at their maximally critical points where the
masses of all linearized perturbations vanish. These theories have been
conjectured to be dual to logarithmic conformal field theories in the
$(d-1)$-dimensional boundary of an AdS solution. We determine the structure of
the linearized perturbations and their boundary fall-off behaviour. The
linearized modes exhibit the expected Jordan block structure and their inner products are
shown to be those of a non-unitary theory. We demonstrate the
existence of consistent unitary truncations of the polycritical gravity theory
at the linearized level for odd rank.
\end{quotation}

\newpage
\pagestyle{plain}

\tableofcontents

\section{Introduction}
\label{sec:introduction}

The perturbative properties of ordinary general relativity in $d=4$ space-time
dimensions can be improved by adding higher derivative terms to the action. The
price one has to pay for rendering the theory renormalizable in this way is
typically the loss of unitarity~\cite{Stelle:1976gc,Stelle:1977ry}.
Recently, specific models in $d\geq 3$ with special choices of higher derivative terms
have attracted renewed attention for several reasons. One is that in $d=3$ they
can provide consistent ghost-free theories of massive
gravitons. This was first observed in the parity violating `topologically massive
theory of gravity' (TMG)~\cite{Deser:1981wh,Deser:1982vy} with three derivatives and more recently
for the parity preserving `new massive gravity'
(NMG)~\cite{Bergshoeff:2009hq} with four derivatives. 
A crucial feature in the construction of NMG is the choice of coefficients in
the four-derivative Lagrangian such that the problematic scalar mode of the
massive graviton becomes pure gauge. Furthermore, there is a critical point
where the mass of the massive graviton vanishes and degenerates with that of the 
massless graviton.

Both features were later extended to higher dimensions by
the discovery of `critical gravity' theories with four derivatives~\cite{Liu:2009bk,Lu:2011zk,Lu:2011ks,Deser:2011xc,Alishahiha:2011yb,Porrati:2011ku}.
At the critical points one typically encounters logarithmic graviton modes that emerge as the replacement for the massive modes. 
These theories importantly have a non-vanishing cosmological constant. Similar
parity preserving theories now also exist in arbitrary dimension and with an
arbitrary (even) number of space-time derivatives and critical points, the
so-called polycritical gravities~\cite{Nutma:2012ss}.
(For other work on massive
gravity
see~\cite{Vainshtein:1972sx,Boulware:1973my,ArkaniHamed:2002sp,deRham:2010ik,deRham:2010kj,Hassan:2011vm,Hinterbichler:2011tt}.)

Another reason for studying polycritical models is provided by the AdS/CFT
correspondence where one would expect a non-unitary logarithmic CFT as the dual
of a polycritical gravity theory~\cite{Lu:2011ks,Hyun:2011ej,Bergshoeff:2012sc} (see
also~\cite{Grumiller:2008qz,Maloney:2009ck}). The non-unitarity of the
logarithmic CFT is related to the fact that the Hamiltonian cannot be
diagonalized on the fields; there is a Jordan
structure~\cite{Gurarie:1993xq,Flohr:2001zs}.
However, the precise structure of the two-point correlation
functions suggests the existence of unitary truncations, and by AdS/CFT also in the
gravity theory~\cite{Bergshoeff:2012sc}. The example of six-derivative
gravity in $d=3$ was treated recently in~\cite{Bergshoeff:2012ev} whereas
four-derivative critical gravity in $d=4$ appeared
in~\cite{Johansson:2012fs}.

To explore this question further, the present paper analyzes the structure of
the various gravitational modes in polycritical gravity in space-time dimensions $d\geq 3$ at the linear level.
We find that an inner product can be defined that reproduces the  
structure expected from logarithmic CFTs. The linearized graviton excitations
around an AdS background can be organized into a hierarchy of higher and higher logarithmic 
dependence near the boundary of AdS. The lowest mode is the usual Einstein
mode, the next one has an additional logarithmic dependence on the AdS
radius, the next one contains $\log^2$ terms and so on. This allows us to
truncate the linearized theory by imposing appropriate boundary conditions on the graviton
fall-off behaviour. A suitable truncation then renders the inner product matrix
between the various modes positive semi-definite. The null states can also be
factored out, but the resulting theory is quite different depending on the rank
of the polycritical gravity theory. The rank is defined as half the
maximum number of space-time derivatives. When the rank is odd, one  arrives at
a unitary model of a single graviton mode. By contrast, the theory becomes
trivial for even rank; the surviving mode has zero energy. This confirms a
conjecture of~\cite{Bergshoeff:2012sc}.
An alternative description of this truncation can be given by defining a
hierarchy of (conserved) charges and then restricting to a superselection sector
in this charge hierarchy. 

While this paper was being completed, the preprint~\cite{Apolo:2012vv}
appeared that discusses the specific case of non-linear critical gravity of
rank 3 in $d=3$ and $d=4$ with the result that truncations that appear to be
unitary at the linearized level may be inconsistent at the non-linear level,
i.e., the truncation is flawed by a linearization instability.
The argument given there seems to extend to the general case independently of how
the linearized theory is completed and this would suggest that our unitary
subsectors exist only in the linearized approximation.

Our paper is structured as follows. In \autoref{sec:quad_theory}, we give
the Lagrangian of the polycritical theory around AdS space whose various
modes will be obtained in \autoref{sec:solutions}.
Then in \autoref{sec:innerproduct} we define and compute the inner product for
these modes. Using either the hierarchy of charges established
in \autoref{sec:charges} or appropriate boundary conditions, we will be able
to define a unitary truncation of the polycritical model in
\autoref{sec:unitarytrunc}. An appendix shows that our inner product is
equivalent to one derived canonically from a two-derivative master action.

\section{Quadratic Lagrangian}
\label{sec:quad_theory}

In this section we briefly review the quadratic Lagrangian around AdS space of
polycritical models of arbitrary rank. But before doing so, it is useful to
first go over the rank one (i.e.~two derivative) case: Einstein gravity with a
cosmological constant.

\subsection{Rank one: Einstein gravity}

Recall that for Einstein gravity with a cosmological constant, we have
the Lagrangian
\begin{equation}
\label{eq:EH_lagrangian}
	\mathcal{L} = \sqrt{-g} ( R  - 2 \Lambda ) .
\end{equation}
The equations of motion state that the cosmological Einstein tensor (that is,
the Einstein tensor plus a term proportional to the cosmological constant)
vanishes,
\begin{equation}
\label{eq:eh_eom}
	G^\Lambda_{\mu\nu} 
		= G_{\mu\nu} + \Lambda g_{\mu\nu}
		= 0 .
\end{equation}
We will perform perturbations around solutions of the equations of
motion as follows,
\begin{equation}
	g_{\mu\nu} = \bar g_{\mu\nu} + g^L_{\mu\nu} = \bar g_{\mu\nu} + h_{\mu\nu} .
\end{equation}
The bar indicates the background solution, and the superscript $L$ the
linear perturbations around it. Thus the linear perturbation of the metric is
given by $h_{\mu\nu}$.
We take the background solution to be an AdS space, which means that the
curvature tensors satisfy
\begin{subequations}
\label{eq:adsbackground}
\begin{align}
	\bar R_{\mu\nu\rho\sigma}  
		& = \frac{ 2 \Lambda}{(d-2)(d-1)} \left( 
			\bar g_{\mu\rho} \bar g_{\nu\sigma} -
			\bar g_{\mu\sigma} \bar g_{\nu\rho} 
		\right) , \label{eq:adsbackground_riemann} \\
	\bar R_{\mu\nu}	
		& =	\frac{ 2 }{(d-2)} \Lambda \bar g_{\mu\nu} , \\
	\bar R 
		& = \frac{ 2 d }{(d-2)} \Lambda , \\
	\bar G_{\mu\nu} 
		& = - \Lambda \bar g_{\mu\nu} \label{eq:einstein_ads} ,
\end{align}
\end{subequations}
with $d$ being the number of space-time dimensions and $\Lambda<0$. 
Instead of the cosmological constant, we can also use the AdS length $\ell$ as
a measure for the background curvature. The two
are related via
\begin{equation}
	\frac{1}{\ell^2} = - \frac{ 2 \Lambda}{(d-2)(d-1)} .
\end{equation}
Note that \eqref{eq:einstein_ads} indeed solves the equations of motion
\eqref{eq:eh_eom}. On this background, the linear equations of motion become
\begin{equation}
\label{eq:eh_eom_lin}
	\left( G^\Lambda_{\mu\nu} \right)^L  = 		
		R^{L}_{\mu\nu} 
		- \frac{2 \Lambda}{(d-2)} h_{\mu\nu} 
		- \frac12 \bar{g}_{\mu\nu} R^L  = 0 ,
\end{equation}
with
\begin{subequations}
\begin{align}
	R^L	& =  
		\bcd_\rho \bcd_\sigma h^{\rho\sigma} - \bar\dal h - \frac{2}{d-2} \Lambda h
		, \label{eq:lin_ricci} \\
	R^L_{\mu\nu} & = 
		\bcd_\rho \bcd_{(\mu} h_{\nu)}{}^\rho  - \frac{1}{2} \bar\dal
		h_{\mu\nu}  - \frac{1}{2} \bcd_\mu \bcd_\nu h  .
\end{align}
\end{subequations}
Taking the trace of the linear equation of motion \eqref{eq:eh_eom_lin} is the
same as linearizing the trace of the non-linear equation of motion
\eqref{eq:eh_eom}, because the cosmological Einstein tensor vanishes by
construction on the background. Either way, we find
\begin{equation}
	\bar g^{\mu\nu} \left( G^\Lambda_{\mu\nu} \right)^L 
		= \left( g^{\mu\nu} G^\Lambda_{\mu\nu} \right)^L
		= \left( 1 - \frac{d}{2} \right ) R^L = 0 .
\end{equation}
Furthermore, the linear equations of motion \eqref{eq:eh_eom_lin} have a gauge
invariance that stems from the diffeomorphism invariance of the non-linear
theory. To be precise, they are invariant under the gauge transformation
\begin{equation}
	h_{\mu\nu} \rightarrow h'_{\mu\nu} 
		= h_{\mu\nu} +  \bcd_{(\mu} v_{\nu)} ,
\end{equation}
for any vector $v_\mu$. This gauge invariance, combined with
the on-shell vanishing of the linearized Ricci scalar $R^L$, implies
\cite{Wald:1984rg} that we can go to the so-called `transverse traceless' gauge,
\begin{align}
	\bcd^\mu h_{\mu\nu} & = 0 , \\
	h & = 0 .
\end{align}
This gauge eliminates the scalar mode (that would otherwise
be a ghost) of $h_{\mu\nu}$, making it a proper spin-2 field.\footnote{As is clear from equation \eqref{eq:lin_ricci}, the transverse
traceless gauge `gauges away' the linear Ricci scalar. This might seem
counter-intuitive, as the linear Ricci scalar is gauge-invariant. But we must
not forget that we used the fact that it vanishes on-shell in order to arrive at
the above gauge. Thus in using this gauge one automatically goes (partially) on-shell.}

In the transverse traceless gauge, the linearized equation of motion
\eqref{eq:eh_eom_lin} simplifies considerably to
\begin{equation}
	\left( G^\Lambda_{\mu\nu} \right)^L = 
		- \frac{1}{2} \left( \bar\dal + 2\ell^{-2} \right)
		h_{\mu\nu} 
		= 0.
\end{equation}
The term $2\ell^{-2}$ may look like a mass term, but it is not. Mass terms in
general break gauge invariance, but the linearized equations of motion were in
fact gauge invariant. Instead, if one were to introduce a mass for the spin-2
field, its equation of motion would read $\left(\bar\dal + 2\ell^{-2} -
m^2\right) h_{\mu\nu} = 0$, with $m$ being the proper mass parameter.

Lastly, the linear equations of motion \eqref{eq:eh_eom_lin} can also be
obtained from the quadratic perturbation of the Lagrangian
\eqref{eq:EH_lagrangian}, which, after partial integration, reads
\begin{equation}
\label{eq:eh_quad_lagrangian}
	\mathcal{L}_2 =  -\frac 1 2 \sqrt{-\bg} \, h^{\mu\nu} \left( G^\Lambda_{\mu\nu}
	\right)^L .
\end{equation}
Indeed, upon varying this quadratic action with respect to $h_{\mu\nu}$ we
recover \eqref{eq:eh_eom_lin}.

\subsection{Einstein and Schouten operators}

The fact that the Lagrangian \eqref{eq:eh_quad_lagrangian} is quadratic in
$h_{\mu\nu}$ is obscured as the linear Einstein tensor $\left(
G^\Lambda_{\mu\nu} \right)^L$ also contains $h_{\mu\nu}$. We can make the
quadratic dependence a bit more transparent by introducing the so-called
Einstein operator $\OG$, upon which the Lagrangian reads
\begin{equation}
\label{eq:eh_lagrangian_quad_op}
	\mathcal{L}_2 =  -\frac 1 2 \sqrt{-\bg} \, h^{\mu\nu} \OG h_{\mu\nu}
	.
\end{equation}
The (cosmological) Einstein operator $\OG$ is defined as
\begin{equation}
	\OG h_{\mu\nu} \equiv \left( G^\Lambda_{\mu\nu} \right)^L .
\end{equation}
Here and in the following we have suppressed the indices on $\OG$. But it is in
fact a tensorial operator, so when we write $\OG h_{\mu\nu}$ we implicitly
mean $\OG_{\mu\nu}{}^{\rho\sigma} h_{\rho\sigma}$. Reading off from equation
\eqref{eq:eh_eom_lin}, the explicit form of the Einstein operator is
\begin{align}
	\OG_{\mu\nu}{}^{\rho\sigma} =
		\bcd^\rho \bcd_{(\mu} \delta_{\nu)}^\sigma
	&	- \tfrac 1 2 \bar\dal \delta_\mu^\rho \delta_\nu^\sigma
		- \tfrac 1 2 \bcd_\mu \bcd_\nu \bg^{\rho\sigma}
		- \tfrac 1 2 \bg_{\mu\nu} \bcd^\rho \bcd^\sigma \nonumber \\
	&	+ \tfrac 1 2 \bg_{\mu\nu} \bar\dal \bg^{\rho\sigma}
		- \frac{2 \Lambda}{d-2} \delta_\mu^\rho \delta_\nu^\sigma
		+ \frac{\Lambda}{d-2} \bg_{\mu\nu} \bg^{\rho\sigma} .
\end{align} 
The Einstein operator has a number of nice properties:
\begin{enumerate}
  \item It is self-adjoint under partial integration:
  		\begin{equation}
  			A^{\mu\nu} ( \OG B_{\mu\nu} ) 
  				= (\OG A_{\mu\nu}) B^{\mu\nu} 
  				+ \textrm{total	derivative}.
  		\end{equation}
  \item It is conserved:
  		\begin{equation}
  			\bcd^\mu \OG A_{\mu\nu} = 0 .
  		\end{equation}
  \item It is gauge invariant:
  		\begin{equation}
  			\OG \bigl[ A_{\mu\nu} + \bcd_{(\mu} v_{\nu)} \bigr] = \OG A_{\mu\nu} .
  		\end{equation}
\end{enumerate}
Here the symmetric $A_{\mu\nu}$, $B_{\mu\nu}$, and $v_\mu$ are completely
arbitrary.

In the following we will also need another operator, the so-called
(cosmological) Schouten operator $\OS$ \cite{Nutma:2012ss}. It is defined
similarly as the Einstein operator, the difference being that it yields the linearized cosmological Schouten tensor when
applied to $h_{\mu\nu}$:
\begin{equation}
	\OS h_{\mu\nu} \equiv \left( S^\Lambda_{\mu\nu} \right)^L .
\end{equation}
In turn, the cosmological Schouten tensor is the usual Schouten
tensor\footnote{Strictly speaking, the usual Schouten tensor has an
additional overall $\tfrac{1}{d-2}$ factor. However, for our purposes it is
more convenient to normalize slightly differently.} plus a term proportional to
the cosmological constant:
\begin{align}
	S^\Lambda_{\mu\nu} 
		& = S_{\mu\nu} - \frac{\Lambda}{d-1} g_{\mu\nu} \nonumber \\
		& = R_{\mu\nu} - \frac{1}{2(d-1)} g_{\mu\nu} R - \frac{\Lambda}{d-1}
		g_{\mu\nu}.
\end{align}
The extra term proportional to the cosmological constant is chosen such that the
cosmological Schouten tensor vanishes on AdS backgrounds,
\begin{equation}
	\bar{S}^\Lambda_{\mu\nu} = 0 .
\end{equation}
The linearized cosmological Schouten tensor reads
\begin{equation}
	\left( S^\Lambda_{\mu\nu} \right)^L
		= R^L_{\mu\nu} 
			- \frac{2 \Lambda}{(d-2)} h_{\mu\nu} 
			- \frac1{2(d-1)} \bar{g}_{\mu\nu} R^L .
\end{equation}
Note that it differs from the linearized cosmological Einstein
tensor \eqref{eq:eh_eom_lin} by a factor of $R^L$:
\begin{equation}
	\left( S^\Lambda_{\mu\nu} \right)^L 
		= \left( G^\Lambda_{\mu\nu} \right)^L
		+ \frac{1}{2} \frac{d-2}{d-1} \bg_{\mu\nu} R^L .
\end{equation}
The Schouten operator on its own does not have striking properties: it is not
self-adjoint, nor is it conserved. However, in combination with the Einstein
operator, things become more interesting:
\begin{enumerate}
  \item $\OG\OS$ is self-adjoint under partial integration:
  		\begin{equation}
  			A^{\mu\nu} (\OG \OS B_{\mu\nu} ) 
  				= (\OG \OS A_{\mu\nu}) B^{\mu\nu} 
  				+ \textrm{total	derivative}.
  		\end{equation}
  		And because $\OG$ on its own is also self-adjoint, $\OG\OS^k$ (the
  		$k$-fold application of $\OS$ followed by $\OG$) is so too.
  \item $\OS$ can be traded for a cosmological constant when
  taking the trace:
  		\begin{equation}
  		\label{eq:gs_trace}
  			\bg^{\mu\nu} \OG \OS A_{\mu\nu} 
  				= \frac{d-2}{2\ell^2} \bg^{\mu\nu} \OG A_{\mu\nu}.
  		\end{equation}
  \item $\OS$ is gauge invariant:
  		\begin{equation}
  			\OS \bigl[ A_{\mu\nu} + \bcd_{(\mu} v_{\nu)} \bigr] = \OS A_{\mu\nu} .
  		\end{equation}
	\item For a symmetric, transverse and traceless tensor (say $C_{\mu\nu}$),
	$\OG$ and $\OS$ are the same:
		\begin{equation}
			\OS C_{\mu\nu} = \OG C_{\mu\nu} =  
			- \frac{1}{2} \left( \bar\dal + 2\ell^{-2} \right)
			C_{\mu\nu} . 
		\end{equation}
\end{enumerate}
The first two properties are crucial for constructing a quadratic theory of
general rank, which we will do now.

\subsection{General rank}

The rank $r$ polycritical Lagrangian around an AdS background
is given by
\begin{equation}
\label{eq:polycritical_lag}
	\mathcal{L}^{(r)}_2 = - \frac{1}{2\tau} \sqrt{-\bg} \, 
		h^{\mu\nu} \OG \prod_{i=1}^{r-1} 
			\left( 2 \OS + m_i^2 \right) 
		h_{\mu\nu} ,
\end{equation}
with $\tau = \prod_{i=1}^{r-1} \left( m^2_i + \frac{d-2}{\ell^2} \right)$. For
rank one, it reduces to the quadratic Einstein Lagrangian
\eqref{eq:eh_lagrangian_quad_op}, as required. 

The non-linear completion for rank one is unique \cite{Deser:1969wk}; it is
simply the Einstein-Hilbert Lagrangian \eqref{eq:EH_lagrangian}. For rank two in
$d =3$ \cite{Bergshoeff:2009hq} or $d=4$ \cite{Lu:2011zk} the non-linear
completion is also unique, because the number of independent curvature
invariants is sufficiently small in those cases. However, for higher rank the
quadratic theory no longer uniquely fixes the non-linear theory, due to the growth of
curvature invariants. One can still
find some non-linear Lagrangian that reproduces the above theory
\eqref{eq:polycritical_lag} for quadratic perturbations. For $d \geq 4$ and
arbitrary rank this was done in \cite{Nutma:2012ss}, while
\cite{Bergshoeff:2012ev} has a non-linear action for $r=3$, $d=3$.
However, finding a unitary interacting theory is not so easy
\cite{Stelle:1976gc,Stelle:1977ry}. We will content ourselves with
knowing one can always write down a non-linear completion.

Since $\OG\OS^k$ is self-adjoint, the equations
of motion that follow from \eqref{eq:polycritical_lag} are simply
\begin{equation}
	\frac{1}{\tau} \OG \prod_{i=1}^{r-1} 
			\left( 2 \OS + m_i^2 \right) 
	h_{\mu\nu} = 0 .
\end{equation}
Upon taking the trace of this, we find with the help of \eqref{eq:gs_trace},
\begin{equation}
	\frac{1}{\tau} \bg^{\mu\nu} \OG \prod_{i=1}^{r-1} 
			\left( 2 \OS + m_i^2 \right) 
	h_{\mu\nu}
		= \bg^{\mu\nu} \OG h_{\mu\nu}
		= \left(1- \frac{d}{2}\right) R^L
		= 0 .
\end{equation}
Note that the use of Schouten operators is crucial in order for the
trace to reduce to the linear Ricci scalar. If one were to use
only Einstein operators in the action \eqref{eq:polycritical_lag}, the trace of
the equations of motion would not be equal to the linear Ricci scalar.

Similarly as in the rank one case, the on-shell vanishing of the linear Ricci
scalar allows us to go to the transverse and traceless gauge. The equations of
motion then become
\begin{subequations}
\label{eq:eom_poly_all}
\begin{align}
	\bcd^\mu h_{\mu\nu} & = 0 , \\ 
	h & = 0 , \\[-12pt]
	\prod_{i=0}^{r-1} \left( \bar\dal + 2\ell^{-2} - m_i^2 \right) h_{\mu\nu} &
	= 0 , \label{eq:eom_polycrit_masses}
\end{align}
\end{subequations}
with $m_0 = 0$. The theory thus contains $r$
propagating gravitons $h^{[i]}_{\mu\nu}$, one of which is always massless while
the others can be massive. Such a graviton mode is a solution to a different
equation of motion than \eqref{eq:eom_poly_all}. Instead it is annihilated by a single
factor of the product of \eqref{eq:eom_polycrit_masses},
\begin{equation}
\label{eq:massdef}
	h^{[i]}_{\mu\nu} : \quad \left( \bar\dal + 2\ell^{-2} - m_i^2 \right)
	h^{[i]}_{\mu\nu} = 0 ,
\end{equation}
while of course still being transverse and traceless. Their on-shell
quasilocal energies can be computed by taking an appropriate integral of
the effective stress-energy tensor, as we will explain in more detail in
\autoref{sec:energy_rankone}. For the modes $h^{[i]}$ defined above the result
is
\cite{Nutma:2012ss}
\begin{equation}
	E^{[i]} 
		= \frac{E_0}{\tau}
			\prod_{\substack{j=0\\j\neq i}}^{r-1}\left( m^2_j - m^2_i \right) .
\end{equation}
Here $E_0$ is the energy of the massless graviton for rank one, that is, the
usual graviton energy in Einstein gravity. If we arrange the masses by size,
\begin{equation}
\label{eq:mass_arrange}
	m^2_1 < \ldots < m^2_{i-1} < m^2_i < m^2_{i+1} < \ldots < m^2_{r-1} ,
\end{equation}
the sign of the energies alternates:
\begin{equation}
	\sgn\left(E^{[i]}_r\right) = - \sgn\left(E^{[i+1]}_r\right) .
\end{equation}
So unfortunately, when the
masses are not degenerate some of the gravitons will always be ghosts, no
matter how one chooses the overall sign of the action. A notable
exception is the $d=3$, $r=2$ case, NMG \cite{Bergshoeff:2009hq}. In three
dimensions the massless graviton does not propagate, but the massive one does. One can then choose the
overall sign of the action such that massive graviton has positive energy and is
not a ghost.

However, for generic dimensions and rank, such a thing is not possible. In an
attempt to ameliorate the situation, one can send all the masses to zero,
thereby reaching the \emph{polycritical point}.

\subsection{Polycritical point}

We define the maximally polycritical point
to be the point in parameter
space where all the masses are zero.
The Lagrangian \eqref{eq:polycritical_lag} then reads
\begin{equation}
	\mathcal{L}^{(r)}_2 = - \frac{1}{2} \left( \frac{2\ell^2}{d-2}
	\right)^{r-1} h^{\mu\nu} \OG \OS^{r-1} 
		h_{\mu\nu} ,
\end{equation}
and the equations of motion that follow from it are
\begin{equation}
\label{eq:eom_poly}
	\OG \OS^{r-1} h_{\mu\nu} = 0 .
\end{equation}
We know from before that this allows us to let $h_{\mu\nu}$ be transverse and
traceless. In this gauge, the Schouten and Einstein operator become the same,
and the remaining equations of motion read
\begin{equation}
\label{eq:eom_poly_og}
	\OG^r h_{\mu\nu} = 0 .
\end{equation}
It is worth stressing that the above is not the correct complete equation of
motion; one must take into account that $h_{\mu\nu}$ is already transverse and
traceless. If this was not the case
we would have to go back to \eqref{eq:eom_poly}.

At the polycritical point the $r-1$ massive modes degenerate with the massless
mode into a single mode.
In addition $r-1$ new modes appear, the so-called log modes
$h^{(I)}_{\mu\nu}$ (with $I = 1,\ldots,r-1$).
These log modes satisfy different equations of motion than the massless
mode; they are annihilated by two or more Einstein operators:
\begin{equation}
\label{eq:log_modes}
	h^{(I)}_{\mu\nu} : \quad 
	\begin{array}{r}
	\OG^{I+1} h^{(I)}_{\mu\nu} = 0, \\ 
	\OG^{I}   h^{(I)}_{\mu\nu} \neq 0,
	\end{array}
\end{equation}
with $I = 0, 1, \ldots , r-1$. Note that $h^{(0)}_{\mu\nu}$
is the usual massless graviton. The action of a single Einstein operator on a log mode gives a `lower' log mode,
\begin{equation}
\label{eq:log_modes_norm}
	\OG h^{(I)}_{\mu\nu} = \frac{d-2}{2\ell^2} h^{(I-1)}_{\mu\nu} .
\end{equation}
If we use the convention $h^{(-1)}_{\mu\nu} = 0$ then \eqref{eq:log_modes}
reproduces the equations of motion.

The main aim of this paper is to analyse the inner product of the log
modes and study the unitarity of the linearized theory. Next, we will
examine explicit solutions to the equations of motion of the log modes.

\section{Linearized log modes}
\label{sec:solutions}
In this section we explicitly present modes of the linearized equations of
motion of the polycritical gravities at their maximally polycritical point. We first recall
certain basic facts about constructing such modes from
\cite{Li:2008dq,Bergshoeff:2011ri} and then give some explicit expressions.
A similar analysis has been recently performed by other authors
\cite{Bergshoeff:2011ri,Chen:2011in,Apolo:2012vv}.

To construct log modes, one first constructs massive as well as massless
transverse traceless spin-2 modes in terms of the highest weight representations
of the symmetry group SO$(2,d-1)$ of AdS$_d$ spacetime.  From the highest weight
states all other states are obtained by acting with the negative root generators of the
algebra. Recall that the mass parameter for spin-2 modes is defined as in
\eqref{eq:massdef}, therefore a general transverse traceless massive spin-2 mode
$\psi_{\mu\nu}$ satisfies
\begin{subequations}
\label{eq:TTmodes}
\begin{align}
\bar{g}^{\mu \nu} \psi_{\mu \nu} &= 0, \\  
	\bcd^\mu \psi_{\mu\nu}  &= 0, \\
 \left( \bar\dal + \frac{2}{\ell^2} - m^2 \right) \psi_{\mu\nu}&	= 0 . 
\end{align}
\end{subequations}
In higher derivative theories at the maximally polycritical point the equations
of motion read \eqref{eq:eom_poly_og} in the transverse traceless gauge. At the
critical points log modes emerge that satisfy different equations of motion
\eqref{eq:log_modes}.

It has been previously observed
\cite{Grumiller:2008qz,Bergshoeff:2011ri,Chen:2011in} that the highest weight
log modes are related to the corresponding massless mode.
The relation is through an overall  factor. 
For the log mode $h^{(I)}_{\mu \nu}$ of index $I$,  the factor is
a polynomial of order $I$ in a function $f$.  To
introduce the function $f$, we first introduce global coordinates on AdS$_d$ in
which the metric of AdS$_d$ takes the form
\be
\label{eq:globalAdS}
ds^2 = \ell^2 \left(- \cosh^2 \rho d\tau^2 + d \rho^2 + \sinh^2 \rho\,
d \Omega_{d-2}^2\right),
\ee 
where $d \Omega_{d-2}^2$ is the unit metric on the round $d-2$ sphere. 
In terms of the coordinates $\tau$ and $\rho$ the function $f$ takes the form 
\be
f = -  i \tau - \log \cosh \rho - \tfrac 1 2  \log 2 .
\ee
The
highest weight massless spin-2 mode $\psi_{\mu \nu}$ in general dimension $d$ is constructed in 
\cite{Chen:2011in}. We will present explicit expressions for $\psi_{\mu \nu}$ 
in four-dimensions in \autoref{sec:boundary_condition}. The massless spin-2
mode $\psi_{\mu\nu}$ is exactly the mode $h^{(0)}_{\mu\nu}$ of the preceding section. 

 We observe the
following properties of the function $f$ and of the highest weight massless
spin-2 mode $\psi_{\mu \nu}$ in general dimension $d$,
\begin{subequations}
\begin{align}
\bar\dal  f &= - \frac{(d-1)}{\ell^2}, \\
\bcd_\sigma f \bcd^\sigma f &= \frac{1}{\ell^2}, \\
\bcd_\sigma f \bcd^\sigma
\psi_{\mu \nu} &=  \frac{d-1}{\ell^2} \psi_{\mu \nu}.
\end{align}
\end{subequations}
From these equations it follows that
\begin{equation}
\bar\dal (f^I \psi_{\mu \nu}) = \frac{I (I-1)}{\ell^2} f^{I-2} \psi_{\mu \nu} +
\frac{(d-1)I}{\ell^2} f^{I-1} \psi_{\mu \nu}  - \frac{2}{\ell^2} f^I  \psi_{\mu
\nu},
\end{equation}
where $I = 0, \ldots, r-1$. As a result,
\begin{equation}
\OG (f^I \psi_{\mu \nu}) = -\frac{I (I-1)}{2\ell^2} f^{I-2} \psi_{\mu \nu}
- \frac{(d-1)I}{2\ell^2} f^{I-1}
\psi_{\mu \nu}.
\end{equation}
Using the above equations one can easily find log modes in the basis
\eqref{eq:log_modes_norm}. The first few modes are
\begin{subequations}
\label{eq:log_modes_all}
\begin{align}
h^{(0)}_{\mu \nu}&= \psi_{\mu \nu}, \\
h^{(1)}_{\mu \nu} &= \left[-\frac{d-2}{d-1}f\right] \psi_{\mu \nu},  \\
h^{(2)}_{\mu \nu} &= \left[\frac{(d-2)^2 }{2
(d-1)^2} f^2  -\frac{(d-2)^2 }{(d-1)^3}f \right]\psi_{\mu \nu} ,  \\
h^{(3)}_{\mu \nu} &= \left[-\frac{(d-2)^3}{6 (d-1)^3} f^3
+\frac{(d-2)^3 }{(d-1)^4} f^2  -\frac{2 (d-2)^3 }{(d-1)^5} f \right]
\psi_{\mu \nu}, 
\\
h^{(4)}_{\mu \nu} &= \left[\frac{(d-2)^4 }{24 (d-1)^4} f^4 
-\frac{(d-2)^4 }{2 (d-1)^5} f^3  +\frac{5 (d-2)^4 }{2
(d-1)^6} f^2   -\frac{5 (d-2)^4 }{(d-1)^7} f \right] \psi_{\mu \nu}.
\end{align}
\end{subequations}
To find a general expression for the index $I$ mode, one needs to solve a
combinatorial equation. This equation can perhaps be solved explicitly, but the
resulting expressions will not be illuminating. Instead one can develop a
recursive algorithm to find log modes that solves \eqref{eq:log_modes_norm} to
whatever order one wants. This is how we have obtained \eqref{eq:log_modes_all}.

\section{Quasilocal energies and the inner product}
\label{sec:innerproduct}

In this section we define an inner product on the (log) solutions. The first
step is to define a bilinear norm $\langle \cdot | \cdot \rangle$ by means of
the quasilocal energy of a solution. Once we have such a norm, the inner product
between two states follows readily.
It turns out that the formulas for the generic case involve the inner product of
Einstein-Hilbert gravity. So, as a starting point, let us see how things work
for rank one.

\subsection{Rank one}
\label{sec:energy_rankone}

The on-shell energy of a solution of the equations of motion can be computed
by
\begin{equation}
	E = \int_\Sigma \textrm{d}^{d-1}x \sqrt{-\bg}
	\, n^\mu \bar\xi^\nu T_{\mu\nu}  ,
\end{equation}
where $\bar\xi^\nu$ is a time-like Killing  vector, $n^\mu$ is the normal to the
Cauchy surface that is being integrated over. The effective
stress-energy tensor $T_{\mu\nu}$ is given by varying the action w.r.t.~the
background metric as if it were dynamical \cite{Iyer:1994ys},
\begin{equation}
	T_{\mu\nu} 
		= - \frac{2}{\sqrt{-\bg}}\frac{\delta \mathcal{L}}{\delta \bar{g}^{\mu\nu}} . 
\end{equation}
In computing the on-shell stress-energy tensor, one has to be a bit careful in
first taking the variation and then going on-shell, because the on-shell Lagrangian vanishes. For the Einstein-Hilbert case
\eqref{eq:eh_lagrangian_quad_op}, the on-shell stress-energy tensor can be
written as
\begin{equation}
	T_{\mu\nu}
		 =  h^{\rho\sigma} \frac{\delta \mathcal
		 G}{\delta \bar g^{\mu\nu}}  h_{\rho\sigma}  ,
\end{equation}
where we have used the short-hand notation $\frac{\delta \OG}{\delta \bg^{\mu\nu}}  h_{\rho\sigma}
		 = \frac{\delta}{\delta \bg^{\mu\nu}} \left( \OG  h_{\rho\sigma} \right) .$
There would have been other contributions, such as the variation w.r.t.~the
background metrics that contract $h^{\rho\sigma}$ and $\OG h_{\rho\sigma}$, but
since they are proportional to the equations of motion they vanish on-shell.
The complete form of the energy of a mode $h_{\mu\nu}$ is thus
\begin{equation}
	E(h) =   \int_\Sigma \textrm{d}^{d-1}x \sqrt{-\bg}
	\, n^\mu \bar\xi^\nu h^{\rho\sigma} \frac{\delta \mathcal
		 G}{\delta \bar g^{\mu\nu}}  h_{\rho\sigma} .
\end{equation}
For physical excitations on AdS spaces this yields a real number.
We can use the above expression to define a norm:
\begin{equation}
	\langle h | h \rangle \equiv E(h) .
\end{equation}
We have dropped the indices on the fields in the norm $\langle \cdot | \cdot
\rangle$ in order to simplify notation, but it is to be understood that we mean
the `full' tensor $h_{\mu\nu}$ in the above, not its trace. The norm 
defines an inner product on the space of solutions as follows:
\begin{align}
	\langle h | k \rangle  
	& = 
		 \frac 1 2 \Bigl( 
		 	\langle h + k | h + k \rangle
		 	- \langle h | h \rangle
		 	- \langle k | k \rangle
		 \Bigr) \nonumber \\
	& = \frac 1 2 \Bigl( E(h+k) - E(h) - E(k) \Bigr) \nonumber \\ 
	& = \frac12 \int_\Sigma \textrm{d}^{d-1}x \sqrt{-\bg} \,
		n^\mu \xi^\nu \lp 
		 	h^{\rho\sigma} \frac{\delta\OG}{\delta\bg^{\mu\nu}} k_{\rho\sigma} 
		+ 	k^{\rho\sigma} \frac{\delta\OG}{\delta\bg^{\mu\nu}} h_{\rho\sigma} 
		\rp .
\end{align}
Because the expression $h^{\rho\sigma} \frac{\delta\OG}{\delta\bg^{\mu\nu}}
k_{\rho\sigma}$ is not obviously symmetric in $h$ and $k$, we have kept the
symmetrization in the above formula for clarity.

\subsection{General rank}

For general rank one can define an inner product on the space of solutions in a
similar manner as for Einstein gravity. At the polycritical point, the
effective stress-energy tensor is
\begin{equation}
	T_{\mu\nu}  
	=  \left( \frac{2\ell^2}{d-2}	\right)^{r-1}
		\sum_{i=1}^r \left(
			\OG^{i-1} h^{\rho\sigma} 
			\frac{\delta \OG}{\delta\bg^{\mu\nu}} 
			\OG^{r-i} h_{\rho\sigma} 
		\right).
\end{equation}
Here we have used the fact that when acting on transverse and traceless
tensors, the Einstein and Schouten operator are the same. Furthermore we
made use of their variation w.r.t.~the background metric being identical after
going on-shell:
\begin{equation}
	\frac{\delta \OG}{\delta \bg^{\mu\nu}}  h_{\rho\sigma} 
		\Bigg|_{\bcd^\mu	h_{\mu\nu} = h = 0}
	\; = \; 
	\frac{\delta \OS}{\delta \bg^{\mu\nu}}  h_{\rho\sigma} 
		\Bigg|_{\bcd^\mu	h_{\mu\nu} = h = 0}.
\end{equation}
The norm for rank $r$, $\langle \cdot | \cdot \rangle_r$, then becomes
\begin{equation}
\label{eq:inner_product_rank_r}
\langle h | h \rangle_r = 
	\left( \frac{2\ell^2}{d-2} \right)^{r-1} 
	\sum_{i=1}^r \bigl< \OG^{i-1} h \big\vert \OG^{r-i} h \bigr>_1 ,
\end{equation}
where $\langle \cdot | \cdot \rangle_1$ is the norm of the rank
one (Einstein-Hilbert) case, as calculated in the last section. This leads to
the following inner product on the log modes:
\begin{equation}
	\left< h^{(I)} \Big\vert h^{(J)} \right>_r 
		=  \sum_{i=1}^r \left< h^{(I-i+1)} \Big\vert h^{(J+i-r)} \right>_1 .
\end{equation}
Since $h^{(-1)}_{\mu\nu} = 0$, the inner product between log modes
$h^{(I)}_{\mu\nu}$ and $h^{(J)}_{\mu\nu}$ vanishes if $I + J < r-1$.
Furthermore, we can relate the inner product for rank $r$ to the inner product
at lower rank if $I > J$:
\begin{align}
	\left< h^{(I)} \Big\vert h^{(J)} \right>_r 
		& =  \sum_{i=2}^r \left< h^{(I-i+1)} \Big\vert h^{(J+i-r)} \right>_1
				\nonumber \\
		& =  \sum_{i=1}^{r-1} \left< h^{(I-i)} \Big\vert h^{(J+i-r+1)} \right>_1
				\nonumber \\
		& = \left< h^{(I-1)} \Big\vert h^{(J)} \right>_{r-1}  .
\end{align}
In the first line we used the fact that $h^{(J-r+1)}=0$ for $J<I\leq r-1$,
and in the second we relabeled the summation index. This allows us to
inductively compute the complete inner product matrix for generic rank
starting from rank 1. The only new bit of information at every step is $\left<
h^{(r-1)} \Big\vert h^{(r-1)} \right>_r$, which is the energy of the maximal log
mode at the given rank. We denote this quantity by $E_{r-1}$:
\begin{equation}
E_{r-1} \equiv \left<h^{(r-1)} \Big\vert h^{(r-1)} \right>_r .
\end{equation}
The inner product matrix for generic rank then reads
\begin{equation}
\label{eq:ipM}
	\left< h^{(I)} \Big\vert h^{(J)} \right>_r =
	\begin{pmatrix}
	0 & 0 & 0 & \cdots& 0 & E_0 \\
	0 & 0  &\cdots & 0 & E_0 & E_1 \\
	0 & \vdots  & 0 & E_0 & E_1 & \vdots\\ 
	\vdots  & 0  & E_0 & E_1 & \vdots & E_{r-3} \\
	0 & E_0  & E_1 & \cdots & E_{r-3}  & E_{r-2} \\ 
	E_0 & E_1  &  \cdots &  E_{r-3} & E_{r-2} & E_{r-1} \\
	\end{pmatrix}  
.
\end{equation}
In \autoref{sec:master_action}  we show that the structure of this inner
 product is same as the one derived canonically from a two-derivative master
 action.
Since $E_0$ is the energy of the massless graviton in Einstein gravity, it is a
positive number. In four dimensions $E_1$ was also found to be positive
\cite{Lu:2011zk}. Based on the explicit solutions of the log modes
\eqref{eq:log_modes_all} and the form of the inner product, we expect all norms
to be non-zero. 

The inner product matrix is indefinite. Regardless of the exact values of the
energies $E_I$, one can always find linear combinations whose energies have
opposite sign. One example is
\begin{equation}
	\left<h^{(r-1)} \Big\vert h^{(r-1)} \right>_r = -
	\left<
		h^{(r-1)} - \frac{E_{r-1}}{E_0} h^{(0)}
	\Big\vert 
		h^{(r-1)} - \frac{E_{r-1}}{E_0} h^{(0)}
	\right>_r.
\end{equation}
Thus one of the above modes is a ghost, implying that the untruncated
linear theory is not unitary.

It is possible, however, to truncate some of the log modes such that the
resulting submatrix of \eqref{eq:ipM} is semi-positive definite. After we have
developed some of the necessary machinery in the next section, we
will demonstrate this in \autoref{sec:unitarytrunc}.

\section{Conserved charges}
\label{sec:charges}

We now turn to the construction of conserved charges in these models following
the method of Abbott--Deser for asymptotically AdS spaces~\cite{Abbott:1981ff,Deser:2002rt,Deser:2002jk}.
It will turn out that one can define an extended hierarchy of charges that makes
reference to the hierarchy of graviton modes defined above. 

\subsection{Abbott--Deser charge}

The method of~\cite{Abbott:1981ff,Deser:2002rt,Deser:2002jk} was applied to
polycritical theories in~\cite{Nutma:2012ss} and makes use of the split of the metric into background
and perturbation $g_{\mu\nu}=\bar{g}_{\mu\nu}+h_{\mu\nu}$, where
$h_{\mu\nu}$ does not need to be small. In the formalism, one splits the
equations of motion for $h_{\mu\nu}$ into linear terms and non-linear terms.
The non-linear terms are then interpreted as an effective
energy-momentum tensor $\mathcal{T}_{\mu\nu}$ that, by the equations of motion,
can be equivalently expressed linearly in $h_{\mu\nu}$ on-shell.
$\mathcal{T}_{\mu\nu}$ is then conserved by the linearised equations of motion. 

From $\mathcal{T}_{\mu\nu}$ one can define a conserved current by
\begin{align}
J_\mu= \mathcal{T}_{\mu\nu}\bar{\xi}^\nu
\end{align}
in terms of a time-like Killing vector $\bar{\xi}^\nu$.
Writing this current in terms of a divergence is possible by virtue
of $\bar{\nabla}_\mu J^\mu=0$  and leads to
\begin{align}
\label{eq:ADcurrent}
J^\mu = \bar{\nabla}_\nu\mathcal{F}^{\mu\nu},
\end{align}
where $\mathcal{F}^{\mu\nu}=\mathcal{F}^{[\mu\nu]}$. In our specific case we
have~\cite{Nutma:2012ss}
\begin{align}
\label{eq:ADpotential}
\mathcal{F}_{\mu\nu} =  \mathcal{F}_{\bar{\xi}} S^{r-1} h_{\mu\nu}
\end{align}
where 
\begin{align}
\mathcal{F}_{\bar{\xi}} h_{\mu\nu} = \bar{\xi}^\rho
\bar{\nabla}_{[\mu}h_{\nu]\rho} 
 + \bar{\xi}_{[\mu}\bar{\nabla}_{\nu]} h 
 -\bar{\xi}_{[\mu}\bar{\nabla}^\rho h_{\nu]\rho} 
  + h^\rho{}_{[mu}\bar{\nabla}_{\nu]} \bar{\xi}_\rho 
  +\frac12 h \bar{\nabla}_\mu \bar{\xi}_\nu.
\end{align}
$\mathcal{F}_{\bar{\xi}}$ is an operation that creates an antisymmetric tensor
out of a symmetric one. It is constructed in such a way that its
derivative relates to the Einstein operator by
\begin{align}
\bar{\nabla}^\nu  \mathcal{F}_{\bar{\xi}} A_{\mu\nu} = \bar{\xi}^\nu
\OG A_{\mu\nu},
\end{align}
when acting on any symmetric tensor $A_{\mu\nu}$. This ensures that
\begin{align}
\bar{\nabla}^\nu \mathcal{F}_{\mu\nu} = \bar{\xi}^\nu \OG \OS^{r-1}
h_{\mu\nu},
\end{align}
which vanishes by the linearized equations of motion~\eqref{eq:eom_poly_og}. The
conserved Abbott-Deser (AD) charge is then given (up to normalization) by the
integral at infinity via
\begin{align}
\label{eq:ADcharge}
Q = \int_{S_\infty} \mathcal{F}^{0i} dS_i.
\end{align}


\subsection{Hierarchy of charges}

We can define a more refined object by considering the following generalization
of \eqref{eq:ADpotential}
\begin{align}
\mathcal{F}_{\mu\nu}^{(I)} :=  \mathcal{F}_{\bar{\xi}} S^{I-1}
h_{\mu\nu} \quad \text{for $I=1,\ldots,r$.}
\end{align} 
A generalized current can be defined as the divergence of this antisymmetric
tensor via
\begin{align}
\label{eq:lowercurrent}
J_\mu^{(I)} := \bar{\nabla}^\nu \mathcal{F}_{\mu\nu}^{(I)} = \bar{\xi}^\nu \OG
\OS^{I-1} h_{\mu\nu}.
\end{align}
For $I=r$ this gives the conserved Abbott--Deser current~(\ref{eq:ADcurrent}):
$J^{\mu}_{(r)}\equiv J^{\mu}$. For $I<r$, the divergence of this
current does not vanish in general and hence the current is not conserved on the
full space of solutions of the linearized theory. But we see from the definition
of $J^\mu_{(I)}$ that it is proportional to the equations of motion for a
lower log graviton mode. Explicitly, for a transverse traceless mode
\begin{align}
J_\mu^{(I)} = \bar{\xi}^\nu \OG^I h_{\mu\nu},
\end{align}
since the Einstein and Schouten operators then coincide.
Hence, by virtue of the definition of the various graviton modes in
\eqref{eq:log_modes}, we find that $J^\mu_{(I)}=0$ when evaluated for modes
$h^{(K)}_{\mu\nu}$ with $K<I$. So, if one restricts to the part of the
space of solutions spanned by modes $h^{(K)}_{\mu\nu}$ with $K<I$, the current
$J_{(I)}^\mu$ is conserved and can be used to define a conserved charge $Q^{(I)}$ on that
subspace. Hence, the AD charge $Q$ of (\ref{eq:ADcharge}) is equal to $Q^{(r)}$.

Whether a given `charge'  $Q^{(I)}$ vanishes or not can be calculated explicitly
for the various modes $h_{\mu\nu}^{(I)}$. We find the distribution of charges displayed
in \autoref{tab:charges}. In that table we use ``n.d.'' to indicate when a
certain charge is not well-defined for a given mode. It is easy to see
that the only mode that has a non-vanishing charge $Q^{(I-1)}$ is the
mode $h^{(I)}$, all others have vanishing charge (if it is defined). This means
that the only mode that has non-vanishing AD charge $Q^{(r)}$ is the highest
log-mode $h^{(r-1)}_{\mu\nu}$.

\begin{table}[t]
\begin{center}
\begin{tabular}{c|C{1cm}C{1cm}C{1cm}C{0.6cm}C{1cm}C{1cm}C{1cm}}
\diagbox{charge}{mode} & $h^{(r-1)}$ &
$h^{(r-2)}$& $h^{(r-3)}$ & $\cdots$ & $h^{(2)}$ & $h^{(1)}$ & $h^{(0)}$\\
\hline
&&&&&&&\\[-10pt]
$Q^{(r)}$ \;\;&  1 & 0 & 0 & $\cdots$& 0& 0 & 0 \\
$Q^{(r-1)}$ &  n.d. & 1 & 0 & $\cdots$& 0& 0  & 0 \\
$Q^{(r-2)}$ &  n.d. & n.d. & 1 & $\cdots$& 0 & 0 & 0 \\
$\vdots$&$\vdots$&$\vdots$&$\vdots$&$\ddots$ & $\vdots$ & $\vdots$ &
$\vdots$ \\[2pt] 
$Q^{(3)}$&n.d.&n.d.&n.d.&$\cdots$& 1 & 0& 0 \\
$Q^{(2)}$&n.d.&n.d.&n.d.&$\cdots$& n.d. &1 & 0\\
$Q^{(1)}$&n.d.&n.d.&n.d.&$\cdots$& n.d. & n.d. & 1
\end{tabular}
\end{center}
\caption{\label{tab:charges}\it Charges of the various modes, normalized
conveniently.
Only the top-most charge $Q^{(r)}$ is conserved in the full theory, but when
the most logarithmic modes are truncated by deleting columns from the left,
lower charges also become well-defined. We have used the abbreviation ``n.d.''
to indicate when a certain charge is not well-defined for a given mode.}
\end{table}

\section{Unitary subsectors}
\label{sec:unitarytrunc}

In this section, we discuss possible unitary truncations of the polycritical
theory at the linearized level. As anticipated in~\cite{Bergshoeff:2012sc} we will find a difference
between odd and even rank.

\subsection{Truncation by superselection}

Looking at \autoref{tab:charges} one sees that one can consistently truncate
to superselection sectors by demanding that certain charges vanish.

Starting with the AD-charge $Q^{(r)}=0$ we truncate out the mode
$h^{(r-1)}\sim\log^{r-1}$.
In the linearized approximation this truncated sector is dynamically closed.
Furthermore, the `charge' $Q^{(r-1)}$ becomes a perfectly well-defined conserved
charge in this truncated model. We can repeat the superselection now to further
truncate consistently to $Q^{(r-1)}=0$.

One would like to continue the truncation until one obtains a standard positive
semi-definite inner product matrix from \eqref{eq:ipM}. Every new step in the
truncation corresponds to removing the last row and column of the inner product matrix. For
even rank this leads to removing all the modes $h^{(r/2)},\ldots,h^{(r-1)}$; the
resulting inner product matrix is identically zero and the theory has become
trivial. This was already observed in the rank $r=2$ case (four derivatives)
in~\cite{Lu:2011zk,Deser:2011xc}.

For odd rank, a truncation to a theory with positive semi-definite two-point
functions can be achieved by restricting to the sector $Q^{(r)}= \ldots
=Q^{(\frac{r+1}{2}+1)}=0$. In this case the inner product matrix becomes almost
identically zero except for one standard correlator in the lower right-hand
corner. This is the structure of a theory with many null states that need to be
quotiented out. The resulting theory then is that of a single mode
$h^{(\frac{r-1}{2})}$ (defined up to the definition of lower
log-modes).
This mode has the standard correlator and positive energy. It appears to be the same model as
standard Einstein gravity in the linearized approximation. 

For low rank, the truncated models (before removing unphysical null states)
have the following inner product matrices: 
\begin{align}
\qquad\qquad r=2: &&&
\lp\;
\begin{array}{cc}
	\cline{1-1}
	\multicolumn{1}{|c|}{0}&E_0\\
	\cline{1-1}
	E_0&E_1
\end{array}\rp
& \longrightarrow &&&
\lp\begin{array}{c}
	0
\end{array}\rp
\nn\\[4pt]
r=3:&&& \lp\;
\begin{array}{ccc}
	\cline{1-2}
	\multicolumn{1}{|c}{0}& \multicolumn{1}{c|}{0}&E_0\\
	\multicolumn{1}{|c}{0}& \multicolumn{1}{c|}{E_0}&E_1\\
	\cline{1-2}
	E_0&E_1&E_2
\end{array}\rp
& \longrightarrow &&&
\lp\begin{array}{cc}
	0 & 0 \\
	0 & E_0
\end{array}\rp
\nn\\[4pt]
r=4:&&&
\lp\;
\begin{array}{cccc}
	\cline{1-2}
	\multicolumn{1}{|c}{0}& \multicolumn{1}{c|}{0}&0&E_0\\
	\multicolumn{1}{|c}{0}& \multicolumn{1}{c|}{0}&E_0&E_1\\
	\cline{1-2}
	0&E_0&E_1&E_2\\
	E_0&E_1&E_2&E_3
\end{array}\rp
& \longrightarrow &&&
\lp\begin{array}{cc}
	0 & 0 \\
	0 & 0
\end{array}\rp\qquad\qquad
\end{align}
Factoring out the null states (that decouple)  one is then left with a standard
CFT for odd rank, identical to that of Einstein--Hilbert gravity, but this time
for the $\log^{(r-1)/2}$-mode. For even rank, the theory trivializes completely.

\subsection{Truncation by boundary condition}
\label{sec:boundary_condition}
In this subsection we discuss fall-offs of various modes, and truncation to
a unitary subsector from the point of view of boundary conditions at spatial
infinity. For concreteness we work with explicit coordinates and we write
expression only in four dimensions. We expect our considerations to
apply more generally. Let us introduce global coordinates in  AdS$_4$ in which
the metric of AdS$_4$ takes the form
\be
ds^2 = \ell^2 \left(- \cosh^2 \rho d\tau^2 + d \rho^2 + \sinh^2 \rho
(d\theta^2 + \sin^2 \theta d \phi^2)\right).
\ee
In these coordinates explicit expressions for the various components of the
massless spin-2 highest weight mode are
\cite{Bergshoeff:2011ri}
\begin{subequations}
\label{eq:massless_mode_4d}
\begin{align}
\psi_{\tau \tau}&=  - \psi_{\tau \phi} = \psi_{\phi \phi} = \exp(- 3 i \tau +
2 i \phi) \sin^2\theta (\cosh\rho)^{-3} \sinh^2\rho, \\
\psi_{\tau \rho} &=  - \psi_{\rho \phi} =  i (\sinh\rho)^{-1}
(\cosh\rho)^{-1} \psi_{\tau \tau},\\
\psi_{\tau \theta} &=  - \psi_{\theta \phi} =  i \cot \theta \psi_{\tau \tau},\\
\psi_{\rho \rho} &= - (\sinh\rho)^{-2}
(\cosh\rho)^{-2} \psi_{\tau \tau} \\
\psi_{\rho \theta} &=  - \cot \theta (\sinh\rho)^{-1}
(\cosh\rho)^{-1} \psi_{\tau \tau},\\
\psi_{\theta \theta} &=  - \cot^2 \theta \psi_{\tau \tau}.
\end{align}
\end{subequations}
Physical excitations correspond to real or imaginary part of these modes. 
Let us now introduce another set of global coordinates $r$ and $t$ as 
\be
r = \ell \sinh \rho, \qquad t = \ell \tau.
\ee
In these coordinates the AdS metric takes the form
\be
ds^2 = - \left(1 + \frac{r^2}{\ell^2}\right) dt^2 + \left(1 +
\frac{r^2}{\ell^2}\right)^{-1} dr^2 + r^2 (d\theta^2 + \sin^2 \theta d \phi^2).
\ee
The boundary of AdS space lies at $r \to \infty$, or equivalently, $\rho \to
\infty$.  In these new coordinates it is fairly easy to check that the
mode \eqref{eq:massless_mode_4d} satisfies the Henneaux-Teitelboim boundary
conditions \cite{Henneaux:1985tv}. In particular, we have
\begin{subequations}
\label{eq:falloff_massless_mode_4d}
\begin{align}
\psi_{tt}  \sim \psi_{t\theta} \sim \psi_{t\phi} \sim \psi_{\theta \theta} \sim
\psi_{\theta \phi} \sim  \psi_{\phi \phi} & \sim
\mathcal{O}\left(\frac{1}{r}\right), \\
\psi_{rt}  \sim \psi_{r\theta} \sim \psi_{r\phi} & \sim
\mathcal{O}\left(\frac{1}{r^4}\right), \\
\psi_{rr} & \sim
\mathcal{O}\left(\frac{1}{r^7}\right).
\end{align}
\end{subequations}
In fact, except for the $rr$ component these fall-offs  saturate
the Henneaux-Teitelboim boundary conditions.  This is expected: since we are
working in the transverse traceless gauge we do not expect to reproduce all fall
offs of \cite{Henneaux:1985tv}, but we do expect that the fall-offs to be strong
enough for the linearized mode \eqref{eq:massless_mode_4d} to be contained in
the phase space defined by those boundary conditions. We schematically denote a
mode saturating the Henneaux-Teitelboim boundary conditions as
$\psi_\mathrm{HT}$.

The index $I$ log mode $h^{(I)}_{\mu \nu}$ behaves asymptotically as
\begin{subequations}
\begin{align}
h^{(I)}_{tt}  \sim h^{(I)}_{t\theta} \sim h^{(I)}_{t\phi} \sim h^{(I)}_{\theta
\theta} \sim h^{(I)}_{\theta \phi} \sim  h^{(I)}_{\phi \phi} & \sim
\mathcal{O}\left(\frac{\log^I r}{r}\right), \\
h^{(I)}_{rt}  \sim h^{(I)}_{r\theta} \sim h^{(I)}_{r\phi} & \sim
\mathcal{O}\left(\frac{\log^I r}{r^4}\right),\\
 h^{(I)}_{rr}  & \sim
\mathcal{O}\left(\frac{\log^I r}{r^7}\right).
\end{align}
\end{subequations}
This is simply because the function $f$ behaves asymptotically as $f \sim -\log
r$.
Thus, if for a rank $r$ polycritical theory one imposes boundary conditions such
that,
\be
h \sim \log^{r-2}r \psi_\mathrm{HT},
\ee
then one clearly truncates away the highest logarithmic mode, that is,
$\log^{r-1}r \psi$. One can also choose to impose a stronger boundary condition
to truncate way more logarithmic modes. In this way one can continue truncating
away higher index logarithmic modes until  one arrives at the boundary
conditions where one obtains a standard positive semi-definite inner product matrix~(\ref{eq:ipM}).
This happens for a rank $r$ theory when one removes the highest 
$\lfloor\frac{r}{2}\rfloor$ log modes. This can be done by imposing boundary
conditions
\be
h \sim \log^{\lceil\frac{r}{2}\rceil-1} r \psi_\mathrm{HT}.
\ee
At this stage it becomes quite clear that the discussion about the choice of
boundary conditions exactly parallels the discussion of the previous subsection based on 
the superselection sector. 

Note that in comparison to the corresponding three-dimensional discussion
\cite{Bergshoeff:2012ev}, our four-dimensional discussion of boundary
conditions is rather schematic.
In three-dimensions there are other independent studies checking the consistency
of $\log$- \cite{Grumiller:2008es,Henneaux:2009pw,Maloney:2009ck} and $\log^2$-
\cite{Liu:2009pha} boundary conditions. Similar studies do not yet exist for the four- and
higher-dimensional settings. 


\section{Discussion and conclusions}

In this paper we have analyzed the structure of polycritical gravity of rank $r$ in $d$ space-time dimensions at the linear level. 
We found that the $r$ different graviton modes on an AdS background satisfy a hierarchical structure $h^{(I)}\sim \log^I r$ in terms 
of the AdS radius $r$. Their inner products can be calculated by using
quasilocal energies and the inner product matrix was found to exhibit a specific triangular structure as expected from a
putative dual logarithmic CFT description. In particular, the inner product matrix is indefinite, reflecting the non-unitary 
structure of the theory.

Following the method of Abbott and Deser one can define a conserved charge in these models 
and only the highest logarithmic mode $h^{(r-1)}$ carries a non-vanishing charge. Restricting to a superselection sector 
where this charge vanishes ---or equivalently modes that have faster fall-off
near the boundary--- one can truncate the model. As we showed, this process can
be iterated until one ends up in a truncated model with positive semi-definite inner product matrix. After modding out the null states one is then left with a unitary model of a single graviton mode in the odd rank case. For even rank, the truncated model trivializes completely. In either case, the truncated model is unitary at the linear level.

The correlator of the single remaining mode in the odd rank case is non-trivial
and identical to that of the Einstein mode in usual two-derivative general
relativity theory. This raises the question whether our truncated model is
nothing but a reformulation of standard general relativity, albeit a rather
complicated one. Since one has to impose appropriate boundary conditions on the
graviton modes to implement the truncation, this idea is reminiscent of the
proposal of~\cite{Maldacena:2011mk} to obtain Einstein gravity from a
conformal higher-derivative gravity theory with appropriate boundary conditions.
However, since our truncation will probably not remain unitary and consistent
when embedded in a non-linear theory~\cite{Apolo:2012vv}, it appears to be
impossible that our model is equivalent with general relativity.


\subsection*{Acknowledgements}

We would like to thank E.~A.~Bergshoeff, G.~Comp\`ere, S.~Fredenhagen, S.~de
Haan, M.~K\"ohn and I.~Melnikov for useful discussions.

\appendix

\section{Master action and the other inner product}
\label{sec:master_action}

In this appendix we relate our inner product to the symplectic inner product
derived from a two-derivative action. To this end we first observe that
the equations of motion \eqref{eq:eom_poly} can also be obtained from an
auxiliary field action, which we call the `master action.' It only contains two
derivatives but it has $r-1$ auxiliary fields $k^{(I)\mu\nu}$.
It takes the form
\begin{equation}
\label{eq:master_action}
	\mathcal{L}_\textrm{master} = 
	- \frac{1}{2} A_{IJ} k^{(I)\mu\nu} \OG  k^{(J)}_{\mu\nu}
	+ \frac{1}{2} B_{IJ} \Bigl( k^{(I)\mu\nu} k^{(J)}_{\mu\nu}
	- k^{(I)}k^{(J)}\Bigr) ,
\end{equation}
with $I,J = 0, \ldots , r-1$ and the symmetric matrices $A$ and $B$ are given by
\begin{align}
	A_{IJ} & = \delta_{I+J,r-1} , \\
	B_{IJ} & = \delta_{I+J,r-2} .
\end{align}
The equations of motion for the field $k^{(r-1-I)}_{\mu\nu}$ read
\begin{equation}
	\OG k^{(I)}_{\mu\nu} 
		 = k^{(I-1)}_{\mu\nu} - \bar{g}_{\mu\nu} k^{(I-1)} , 
\end{equation}
from which we have, after taking the trace,
\begin{align}
	k^{(I-1)}_{\mu\nu} 
	& = \OG  k^{(I)}_{\mu\nu} 
		- \frac{1}{d-1} \bar{g}_{\mu\nu} \OG  k^{(I)} \nonumber \\
	& = \OS k^{(I)}_{\mu\nu} .
\end{align}
The complete set of equations of motion become
\begin{subequations}
\label{eq:aux_modes}
\begin{align}
	\OG  k^{(0)}_{\mu\nu} & = 0 , \\
	k^{(0)}_{\mu\nu} & = \OS  k^{(1)}_{\mu\nu} , \\
	k^{(1)}_{\mu\nu} & = \OS  k^{(2)}_{\mu\nu} , \\
	& \;\; \vdots \nonumber \\
	k^{(r-2)}_{\mu\nu} & = \OS k^{(r-1)}_{\mu\nu} .
\end{align}
\end{subequations}
Upon eliminating the $r-1$ auxiliary fields $k^{(I \neq r -1)}_{\mu\nu}$, and
calling $k^{(r -1)}_{\mu \nu}$ to be $h_{\mu \nu}$, we recover the original
equations of motion \eqref{eq:eom_poly}. For the rest of the discussion we take
the $k^{(I)}$s to be transverse and traceless, which follows from the equations
of motion. This allows us to freely replace Schouten operators on
$k^{(I)}$ with Einstein operators.

For the two derivative master action  \eqref{eq:master_action}  the symplectic
inner product can be simply computed following the formalism reviewed in
\cite{Porrati:2011ku}. Up to an over-all normalization, the symplectic inner product for a rank $r$ theory takes
the form
\be
\langle \psi || \phi\rangle_r =  \int_\Sigma \textrm{d}^{d-1}x \sqrt{-\bar g}
\bar{g}^{00} \left(\sum_{I,J=0}^{r-1} A_{IJ} (\psi^{(I)}_{\mu \nu})^* (\bcd_0
\phi^{(J)})^{\mu \nu} \right), \label{eq:symplectic_inner}
\ee
where $\psi^{(I)}$ and $\phi^{(I)}$ are the auxiliary field configurations
associated with the configurations 
$\psi$ and $\phi$ respectively. The integration is done over a constant $\tau$
Cauchy surface $\Sigma$ and the index $0$ denotes the time components of the
various tensors in the coordinates \eqref{eq:globalAdS}. Furthermore, to
distinguish this inner product from that of \autoref{sec:innerproduct} we use the double line notation $\langle \psi||\phi\rangle_r$. The subscript $r$ denotes the rank of the theory. From expression \eqref{eq:symplectic_inner} and from the form of the matrix $A_{IJ}$
it follows that
\begin{align}
\langle \psi||\phi\rangle_r 
&= \int_\Sigma \textrm{d}^{d-1}x  \sqrt{-\bar g} \bar{g}^{00}
\left(\sum_{I=0}^{r-1} (\psi^{(I)}_{\mu \nu})^* (\bcd_0
\phi^{(r-1-I)})^{\mu \nu} \right)\nn\\
&= \int_\Sigma \textrm{d}^{d-1}x  \sqrt{-\bar g} \bar{g}^{00}
\left(\sum_{I=0}^{r-1} (\OG^{r-1+I} \psi^{(r-1)}_{\mu \nu})^* (\bcd_0
\OG^{I} \phi^{(r-1)})^{\mu \nu} \right) \nn\\
&= 
\sum_{I=0}^{r-1} \left< \OG^{r-1+I} \psi \Big\vert\Big\vert
\OG^{I} \phi \right>_1\nn\\
&= 
\sum_{j=1}^{r} \left< \OG^{r-j} \psi \Big\vert\Big\vert
\OG^{j-1} \phi  \right>_1,
\end{align}
where in going from the first to the second line we have used
\eqref{eq:aux_modes}, in going from the second to the third line we have used
the notation  $\langle \psi||\phi\rangle_1$ which denotes the (appropriately
normalized) symplectic inner product for the rank one theory, and finally in
going from the third to the fourth line we have renamed the dummy variable $I$
to $j = I +1$.
From this last equality and equation \eqref{eq:inner_product_rank_r} we immediately see 
that the two inner products give rise to the identical matrix structure over log modes.

\printbibliography

\end{document}